\documentclass[reprint,amsmath,amssymb, aps,longbibliography]{revtex4-2}
\usepackage{graphicx}
\usepackage{dcolumn}
\usepackage{bm}
\usepackage{pdfcomment}
\usepackage{xcolor}
\graphicspath{{./fig/}}
\usepackage{float}

\begin{document}
\newcommand{\gcc}{g\,cm$^{-3}\,$}
\newcommand{\Avion}{{\scshape AvIon}}

\title{Reconciling Ionization Energies and Band Gaps of Warm Dense Matter \\Derived with {\it Ab Initio} Simulations and Average Atom Models}

\author{G. Massacrier}
\email{gerard.massacrier@ens-lyon.fr}
\affiliation{%
Univ Lyon, Univ Lyon1, Ens de Lyon, CNRS, Centre de Recherche Astrophysique de Lyon UMR5574, F-69230, Saint-Genis-Laval, France
}

\author{M. B\"ohme }%
\email{m.boehme@hzdr.de}
\affiliation{%
 Helmholtz-Zentrum  Dresden-Rossendorf,  Bautzner  Landstra{\ss}e 400,  D-01328  Dresden,  Germany
  }%
\affiliation{%
 Technische Universit\"at Dresden, D-01062 Dresden, Germany
}%

\author{J. Vorberger}
\email{j.vorberger@hzdr.de}
\affiliation{%
 Helmholtz-Zentrum  Dresden-Rossendorf,  Bautzner  Landstra{\ss}e 400,  D-01328  Dresden,  Germany
 }%

\author{F. Soubiran}
\affiliation{\'Ecole Normale Sup\'erieure de Lyon, Universit\'e Lyon 1, Laboratoire de G\'eologie de Lyon, CNRS UMR 5276, 69364 Lyon Cedex 07, France}
\affiliation{CEA DAM-DIF, 91297 Arpajon, France}

\author{B. Militzer}
\email{militzer@berkeley.edu}
\affiliation{Department of Earth and Planetary Science, Department of Astronomy, University of California, Berkeley, CA 94720, USA}

\date{\today}

\begin{abstract}
Average atom (AA) models allow one to efficiently compute electronic and optical properties of materials over a wide range of conditions and are often employed to interpret experimental data. However, at high pressure, predictions from AA models have been shown to disagree with results from {\it ab initio} computer simulations. Here we reconcile these deviations by developing an innovative type of AA model, \Avion, that computes the electronic eigenstates with novel boundary conditions within the ion sphere. Bound and free states are derived consistently. 
We drop the common AA image 
that the free-particle spectrum starts at the potential threshold, which we found to be incompatible with {\it ab initio} calculations. We perform {\it ab initio} simulations of crystalline and liquid carbon and aluminum over a wide range of densities and show that the computed band structure is in very good agreement with predictions from \Avion.
\end{abstract}


\maketitle


\section{Introduction}
The electronic and optical properties of materials change drastically with increasing pressure or density, which affects their equation of state, opacity, and transport properties. These changes have profound implications for the interior structure and evolution of stars and planets~\citep{Oswalt1996,Militzer2016}. The modification of electronic states with temperature and pressure is a very active research subject~\cite{graziani2014,Flores-Livas2015}. The characterization of electronic properties is especially challenging in the regime of warm dense matter (WDM) because pressure and temperature are high, particles are strongly interacting, and the system is partially degenerate. One relevant phase transition is the dissociation and metallization of hydrogen~\cite{Wigner1935,Saumon1992,Weir1996,Knudson1455} that changes from an insulating, molecular substance to a monoatomic, electrically conducting fluid. This phase change is a manifestation of a Mott transition~\cite{Mott1968,mottbook}. 

Due to the development of high-power lasers and other facilities, the capabilities to create and diagnose WDM have increased substantially and a variety of new experimental results have been obtained~\citep{Vinko2010,Ciricosta2012,Vinko2012,Hoarty2013,Vinko2015,Ciricosta2016,Kasim2018}. Many characterize modifications of the atomic structure due to the plasma environment. A key quantity is the degree of ionization, which may be enhanced by the continuum lowering phenomenon, or ionization potential depression (IPD), that describe the reduction of the ionization energy compared to the value of an isolated ion. X-ray Thompson scattering measurements (XRTS) provide direct access to the IPD~\citep{Fletcher2013,Fletcher2014,Kraus:2018ab}. The analysis of the scattering signal is based on the Chihara formalism~\citep{Chihara2000,Wunsch2011}, which relies on the separation of bound and free electronic states as well as well-defined direct transitions between the two.
There are experimental results both supporting and contradicting the two classic IPD models of Stewart-Pyatt (SP)~\citep{Stewart1966} and Ecker-Kr\"oll (EK)~\citep{Ecker1963}. SP models are widely used to predict plasma properties like ionization balance, equation of state, opacities, electrical and thermal conductivities. 

There are a variety of theoretical approaches to study WDM  and the IPD~\cite{Blancard:2004aa,Hu:2017ab,Roepcke2019}.
With {\it ab initio} simulations, one follows the evolution of many ions in a periodic simulation cell by coupling density functional theory (DFT) 
\citep{HohenbergKohn1964,Mermin1965} 
to molecular dynamics (MD)~\cite{Car1985}. 
In comparison, average atom (AA) models are computationally far less demanding because they place only a single nucleus at the center of a cloud of electrons and determine the resulting states. 
Many variations of the AA approach exist~\citep{Blancard:2004aa,Wilson:2006aa,Piron:2011by,Starrett:2019aa}. However, when these methods are applied to guide or interpret experiments at extreme densities or highly degenerate plasma conditions, the results appear to deviate from predictions of {\it ab initio} methods~\cite{Driver:2018aa,Hu:2017ab,Iglesias:2018ab,Hu:2018aa}, which provided the initial motivation for this study. 

In this paper, 
we use DFT calculations to study the evolution of electronic states of dense C and Al over three orders of magnitude in density and then reproduce them with a novel, general-purpose type of AA model. For the latter, we adopt an alternate view of the continuum. Usually, AA models embed the ion  
in some kind of jellium and 
the continuum of states starts at the threshold of the potential. We determine that it is this property 
that leads to a disagreement with {\it ab initio} results. As a matter of fact, valence and conduction bands as derived from DFT-MD simulations exhibit no clear link to such a potential threshold.

\begin{figure}[t]
    \includegraphics[width=0.48\textwidth]{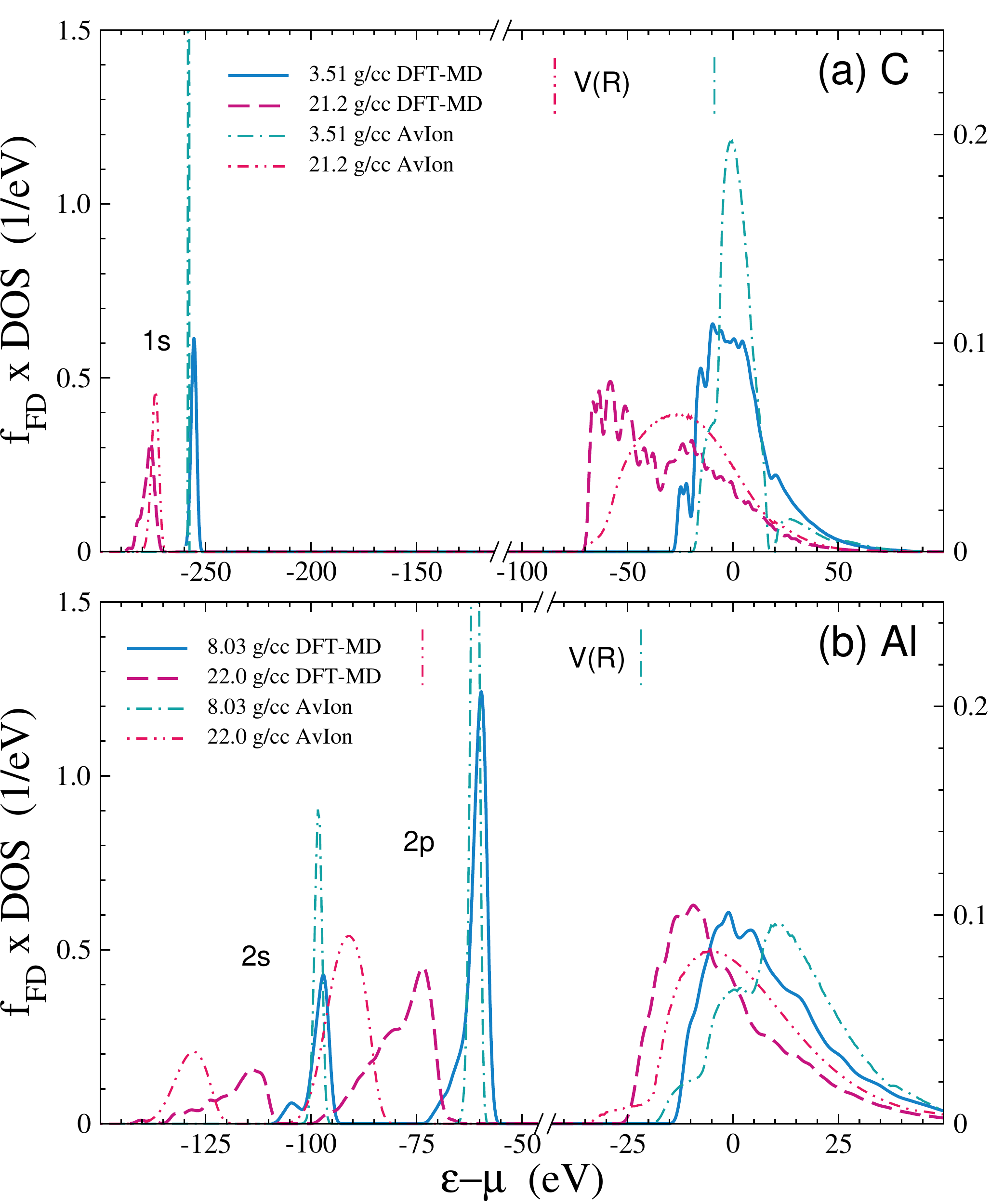}
    \caption{ Density of states (multiplied by the Fermi-Dirac factor) at various densities and $T=12.5~$eV for liquid (a) carbon and (b) aluminum. Energies are counted from the chemical potential. Results are from DFT-MD or \Avion~simulations as shown in the legends. The position of the threshold potential $V(R)$ for each \Avion~simulation is also shown as vertical bars at the top of each panel. For better readability, we introduced different y scales for the valence bands (left y axis) and the continuum states (right y axis).} 
    \label{fig:DOS_ALL}
\end{figure}

\section{Ab initio simulations} 

We performed {\it ab initio} calculations for dense carbon and aluminum with the {\scshape Abinit} code~\cite{Kang2016, Gonze2016, Gonze2009, Bottin2008, Bruneval2006, Torrent2008}. For both materials, we performed static DFT calculation using perfect lattices ($T=0$~eV) and DFT-MD simulations at a finite temperature ($T=12.5$~eV). For aluminum, we employed the local density approximation (LDA) 
to incorporate exchange-correlation (XC) effects~\cite{Perdew1992} 
and the PBE-GGA functional was used for carbon~\cite{Perdew:1996}. For both materials, we generated new pseudopotentials with very small core radii using the Opium code~\cite{OPIUM}. The 1s core electrons were frozen in the norm conserving pseudopotential for aluminium. For carbon, we even performed all-electron calculations. 

As the predicted crystal structures of aluminium and carbon change with density, we change the structure at $T=0$ according to Ref.~\cite{Benedict2014} for carbon and Ref.~\cite{Pickard2010} for aluminium. For carbon, the order of the phases is diamond (3.52 to 7.66 \gcc), BC8 (7.87 \gcc), 
simple cubic (sc, 15.9 \gcc), simple hexagonal (sh, 23.29 \gcc) and face-centered cubic (fcc, 46.36 to 640.83 \gcc).
A $12^3$ k-point grid with a plane wave cutoff energy of $600$~Hartrees (Ha) 
was used for diamond. The BC8, sc, and sh calculations used a k-point grid of $32^3$ points and a cutoff energy of $700 \, $Ha. 
The static $T=0$ calculations for aluminium were carried out for both 
the fcc and body-centered cubic (bcc) \cite{FIQUET2019243} phases 
to get a better understanding on how the phase affects the valence band gap.  For the fcc unit cell a $64^3$ k-point grid with 12 bands and a plane wave cutoff energy of $400 \, $Ha has been used. The bcc calculations used a $12^3$ k-point grid with 20 bands and a plane wave cutoff at $400 \,$Ha.
The band gap reported for these solid phases is the indirect gap over the sampled Brillouin zone.

The MD simulations at $T=12.5$~eV were run using a Nos{\'e}-Hoover thermostat 
and with a time step of $\delta t=0.09$~fs and with 8 atoms in the periodic supercell which was sampled only at the $\Gamma$-point. The number of bands considered was $450$ and the plane wave cutoff was $500$~Ha for carbon. The aluminum DFT-MD simulations considered $630$ to $640$ bands and a cutoff energy of $400$~Ha. 
Electronic density of states (DOS) and band gap results at finite temperatures were derived by averaging over the DOS of $10$ sample configurations obtained via DFT-MD and which where postprocessed with a static run using $8^3$ k-points.  These DOS were averaged after aligning them first at the chemical potential $\mu$ (a few eV below the Fermi energy at these low temperatures).

Figure~\ref{fig:DOS_ALL} shows the DOS, weighted by the Fermi-Dirac factor $1/(1+\mathrm{e}^{(\varepsilon-\mu)/kT})$, at $T=12.5$~eV and for 
densities a few times the solid density. For illustration purposes the DOS were smoothed using a Gaussian with a HWHM of 0.8~eV. For carbon, the peaks below $-250$~eV represent the 1s states while the 2s and 2p states have already merged with the continuum at these conditions. 
With increasing density, we find that the continuum edge shifts towards lower energies when compared with the Fermi energy, consistent with earlier findings~\cite{Driver2015}. For the following discussion, we define the valence band gap (VBG) of carbon to be the gap between the 1s states and the bottom of the continuum band.  

For aluminum at the lowest density of $8.03$~\gcc, the 2s and 2p states form two well-separated peaks. At $22.0$~\gcc, these peaks have broadened. At yet higher densities, they will merge (not shown). The VBG of aluminum is defined to be the gap between the uppermost 2s/2p state and the continuum band.

Our simulations show consistently that the continuum edge shifts towards lower energies, well below the Fermi energy, as density increases. Eventually gaps between valence and continuum bands close. In traditional AA models, the bottom of the continuum 
begins 
at the maximum of the  potential and a state is pressure-ionized as soon as it crosses this threshold. This 
leads to inconsistent results between the DFT and AA approaches because in DFT, one derives the free-particle spectrum without introducing 
such a threshold. 
In order to resolve this discrepancy and to reproduce the fully self-consistent many-body DFT calculations, we developed a novel type of AA approach, \Avion, that is much more efficient than DFT-MD and can be readily applied to arbitrary materials and thermodynamic conditions. As we will show below, \Avion~reproduces the DFT-MD predictions very well. It also explains apparent discrepancies between the usual AA models and DFT-MD simulations. 

\section{The \Avion~model} 

In {\it ab initio} many-body simulations, the band structure is obtained from eigensolutions of the Schr\"odinger equation in a periodic box for an effective potential resulting from a collection of electrons and nuclei, while taking advantage of the Kohn-Sham scheme~\citep{KohnSham1965}. In our \Avion~approach, we intend to solve the same equations to obtain the band structure and wavefunctions around a single nucleus in a limited volume. The only connection with the plasma environment arises from the boundary conditions of the wavefunctions at the surface of that volume. In an actual many particle system these boundary conditions are highly complex and vary from ion to ion depending on the environment. Still these variations can be represented well enough with approximate methods as long as the predictions are validated through comparisons with many-body simulations.

\Avion~is a spherical model with a neutrality radius, $R$. 
To obtain agreement with many-body simulations, we introduce
novel boundary conditions at $R$ 
that allows us to derive bound and free states within a common scheme. As we will show, this has deep implications for the band structure of the continuum states and for the magnitude of band gaps.
Conversely, traditional AA models set the boundary conditions at infinity. This leads to a clear distinction between a bound (discrete) and a free (continuous) spectrum, which is not compatible with predictions of many-body simulations at high density. 

For an element with nuclear charge, $Z$, and at given temperature, $T$, the \Avion~approach self-consistently determines the electronic density profile, $n_e(r)$, and an effective potential, $V(r)$, within the Kohn-Sham scheme~\citep{KohnSham1965}. 
The spherical potential
\begin{equation}
  V(r)=-\frac{Ze^2}{r}+v_{\rm H}(r)+v_{\rm xc}(n_e(r))
  \label{Eq.potential1}
\end{equation}
is the sum of the Coulombic nuclear attraction, the Hartree potential $v_{\rm H}$ resulting from the electronic density through Poisson equation, and an exchange-correlation 
part in the LDA approximation. (We use~\cite{Tanaka1986, Ichimaru1987, Ichimaru1993}.) The potential $v_{\rm H}$ inside $R$ can be written, up to a constant, as a function of the sole charge inside this same volume:
\begin{equation}
  v_{\rm H}(r) = v_{\rm H}(0)-e^2\int_0^r\frac{Q(r')}{r'^2}\,\mathrm{d}r',
  \label{Eq.potential2}
\end{equation}
where  $Q(r)=\int_0^r 4\pi r'^2 n_e(r')\,\mathrm{d}r'$ is the number of electrons inside radius $r$.
We do not need to assume any definite value for $V$, and leave $v_{\rm H}(0)$ unspecified. 

The electronic density (including all electrons) is in turn determined from the spherical eigenstates 
$\phi_{\varepsilon\ell m}=\frac{1}{r}P_{\varepsilon \ell}(r)Y_\ell^m(\hat{\bf{r}})$ 
in the potential $V$. Their radial parts obey
\begin{equation}
 \frac{\mathrm{d}^2P_{\varepsilon\ell}}{\mathrm{d}r^2}=\Big(\frac{\ell(\ell+1)}{r^2}+\frac{2m}{\hbar^2}(V(r)-\varepsilon)\Big)\,P_{\varepsilon\ell}
  \label{Eq.wf}
\end{equation}
(with $m$ the electron mass, $\hbar$ the Planck constant), are normalized to unity inside the ion sphere, 
and populated according to a Fermi-Dirac distribution. Accordingly,
\begin{equation}
  4\pi r^2 n_e(r)=\sum_{\ell}2(2\ell+1)\int{\rm d}\varepsilon \frac{g_\ell(\varepsilon)}{{\rm e}^{(\varepsilon-\mu)/kT}+1}|P_{\varepsilon l}(r)|^2 \;,
  \label{DOS_AvIon}
\end{equation}
where the $g_\ell(\varepsilon)$ are partial DOS for the angular momentum $\ell$.
The chemical potential $\mu$ is adjusted so that $Z$ electrons exactly pile up inside of radius $R$, i.e. $Q(R)=Z$. From electro-neutrality one has $\frac{\mathrm{d}V}{\mathrm{d}r}(R)=0$ (neglecting $v_{\rm xc}$). 
The value $V(R)$ defines the potential threshold on the energy scale. 

The distinctive feature of \Avion~is in the way the DOS is constructed. 
The continuum in traditional AA models starts above that threshold 
(setting $V(r\ge R)=0$), and includes
all $\varepsilon>V(R)$ solutions, matching them to a combination of Bessel functions (free solutions) at the $R$ boundary. We show in this paper that the band structure inherent to condensed matter, which persists for crystals under WDM conditions as shown recently by Bekx {\it et al.}~\cite{Bekx2020} (see in particular their Fig.~7), or in DFT simulations, challenges this simple view, yet can be recovered with a modified AA model.

For bound states, the concept of bands instead of isolated states (which give a $\delta$-function contribution to some $g_\ell(\varepsilon)$) has already been used in a number of AA models~\citep{Rozsnyai:1972aa,Massacrier:1994aa,Potekhin:2005aa,Massacrier:2011aa}. Regarding states above threshold, some preliminary works have explored the modification of the DOS due to neighbors using multiple scattering theory~\citep{Wilson:2011yg,Starrett:2018ab,Starrett2020}. While this is in the spirit of our work, our aim is to avoid the huge complexity it adds.

In \Avion~for every angular momentum, $\ell$, the partial DOS $g_\ell(\varepsilon)=\sum_k g_{k\ell}(\varepsilon)$ is a 
sum over successive energy bands $k$. 
For given $\ell$, all wavefunctions within the $k$-th band have $k-1$ nodes in the interval $]0,R[$. 
In addition, the lower $\varepsilon_{k\ell}^{-}$ and upper $\varepsilon_{k\ell}^{+}$ energy limits of  this  $k$-th band result from the 
eigensolutions of the radial Schr\"odinger equation (Eq.~(\ref{Eq.wf}))
with  the respective boundary conditions $\frac{{\rm d}(P_{\varepsilon\ell}/r)}{{\rm d}r}(R)=0$ and $P_{\varepsilon\ell}(R)=0$.
With these two conditions, we intend to mimic the bonding and antibonding character of typical molecular wavefunctions. 
Note that the $P_{\varepsilon\ell}(R)=0$ condition for the upper band limits does not imply 
that there is a hard wall at radius $R$, but that the total (over the whole plasma) wavefunction, a part of which we solve for in the ion sphere, has a node at $R$. We emphasize that we no longer make a distinction between bound (below threshold) and free (above threshold) states, as is done in traditional AA models. Still we need to make an assumption for the DOS inside each band. 

Between the respective limits $\varepsilon_{k\ell}^{\pm}$, we choose in this work the Hubbard functional form $g_{k\ell}(\varepsilon)=\frac{2}{\pi\delta^2}[(\varepsilon_{k\ell}^{+}-\varepsilon)(\varepsilon-\varepsilon_{k\ell}^{-})]^{1/2}$, where $\delta=\frac{1}{2}(\varepsilon_{k\ell}^{+}-\varepsilon_{k\ell}^{-})$ \cite{Hubbard:1964aa}. This is undoubtedly oversimplified. The DOS of the bands is actually controlled by a complex interplay between the ion and its neighbours, and its computation requires involved techniques. For instance the recent work of 
Starrett \& Shaffer~\citep{Starrett2020} 
uses multiple scattering approach, but requires around one hour of CPU time. Our goal is precisely to bypass this computationally highly demanding step: 
\Avion~necessitates a few seconds per point~\footnote{on a laptop equipped with a quad-core 2.5~GHz Intel Core I7 and 16~Gb of RAM.}.
With this DOS at hand and wavefunctions computed for a set of energies inside the bands, 
the electronic density can be obtained from Eq.~(\ref{DOS_AvIon}).

As a summary, the three parameters $(Z,T,R)$ entirely determine an \Avion~calculation. The iterative procedure is initiated with 
a trial electronic density as input. The potential is determined through Eqs.~(\ref{Eq.potential1}) and (\ref{Eq.potential2}). Bands and wavefunctions are obtained from the solutions of Eq.~(\ref{Eq.wf}). The resulting electronic density is constructed from Eq.(\ref{DOS_AvIon}) with $\mu$ adjusted for electroneutrality. The result is compared with the input.  If different,  a new loop is initiated with a mix of the two densities, and the procedure is continued until full convergence is achieved.

To compare with results at given ion density $n_i$ (or mass density $\rho=Am_{\mathrm u}n_i$, where $A$ is the atomic mass of the element), we set in this paper $R$ equal to the Wigner-Seitz radius, $R_{\mathrm WS}=(3/(4\pi n_i))^{1/3}$. Note that this is only one plausible choice among other possibilities that we don't explore in this work~\footnote{$R^3n_i$ could 
be given an other value, for instance if 
  $R$ is rather viewed as half the distance to the nearest neighbouring ion. This is clear for a crystal structure, but see also e.g. Fig.~2 in \cite{Simoni:2020aa} or Fig.~4 in \cite{Vorberger:2020aa} which compare different radii. Energy bands can also be broadened by letting $R$ fluctuate (liquid case), for instance according to some ion-ion pair correlation function. In this paper we put aside these refinements.\label{footnoteradii} }.

\begin{figure*}[t]
    \centering
    \includegraphics[width=0.49\textwidth]{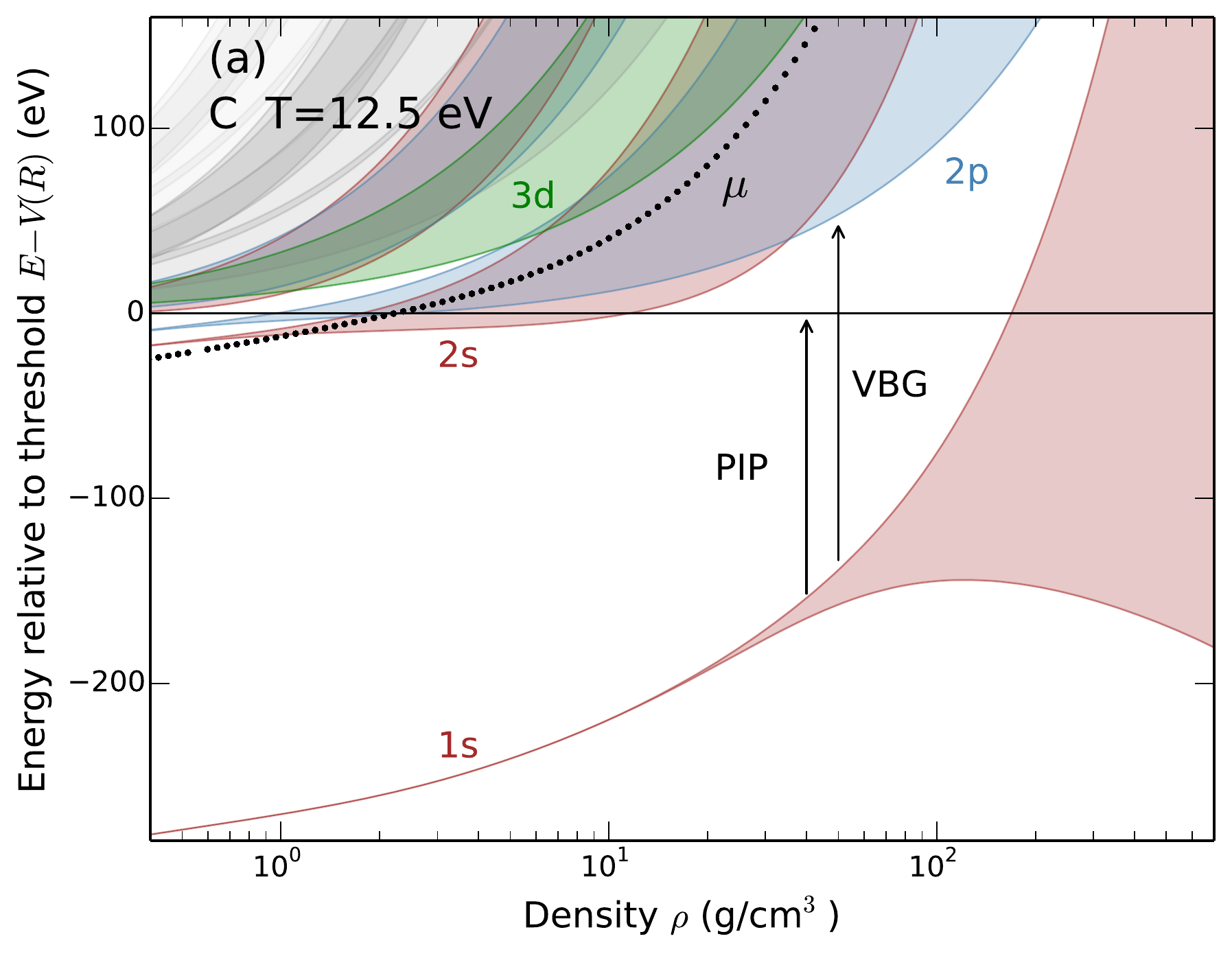} 
    \includegraphics[width=0.49\textwidth]{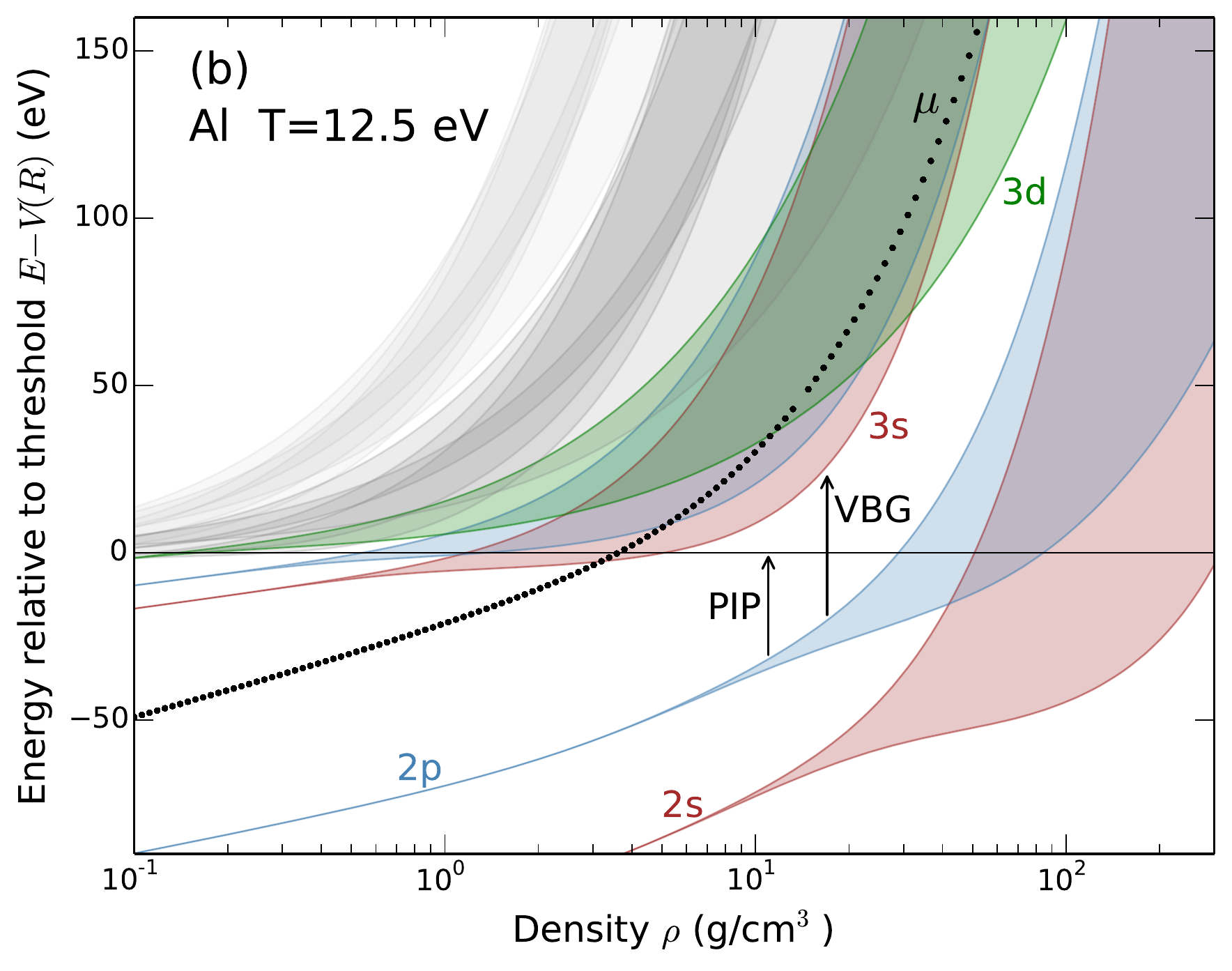} 
    \includegraphics[width=0.49\textwidth]{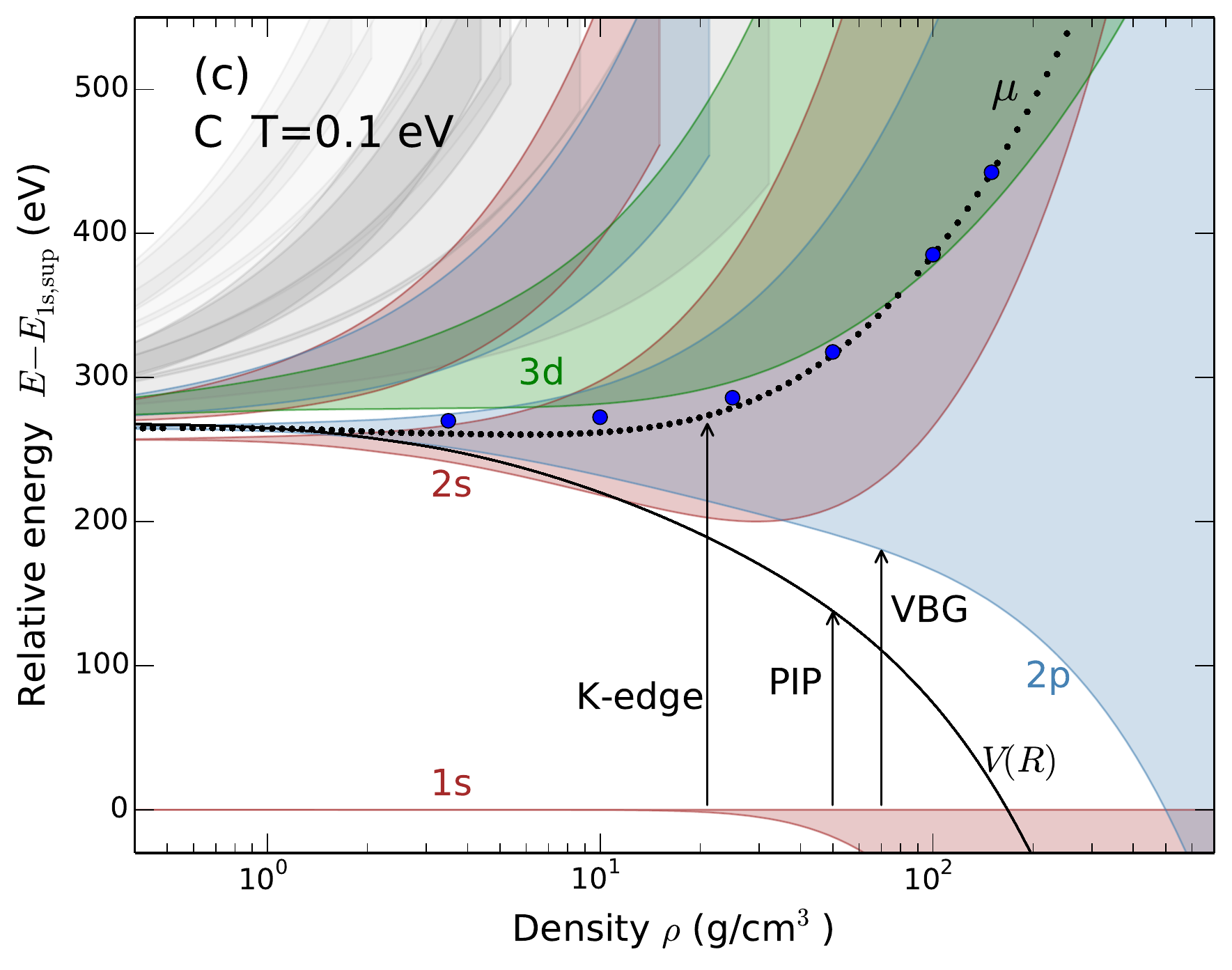}
    \includegraphics[width=0.49\textwidth]{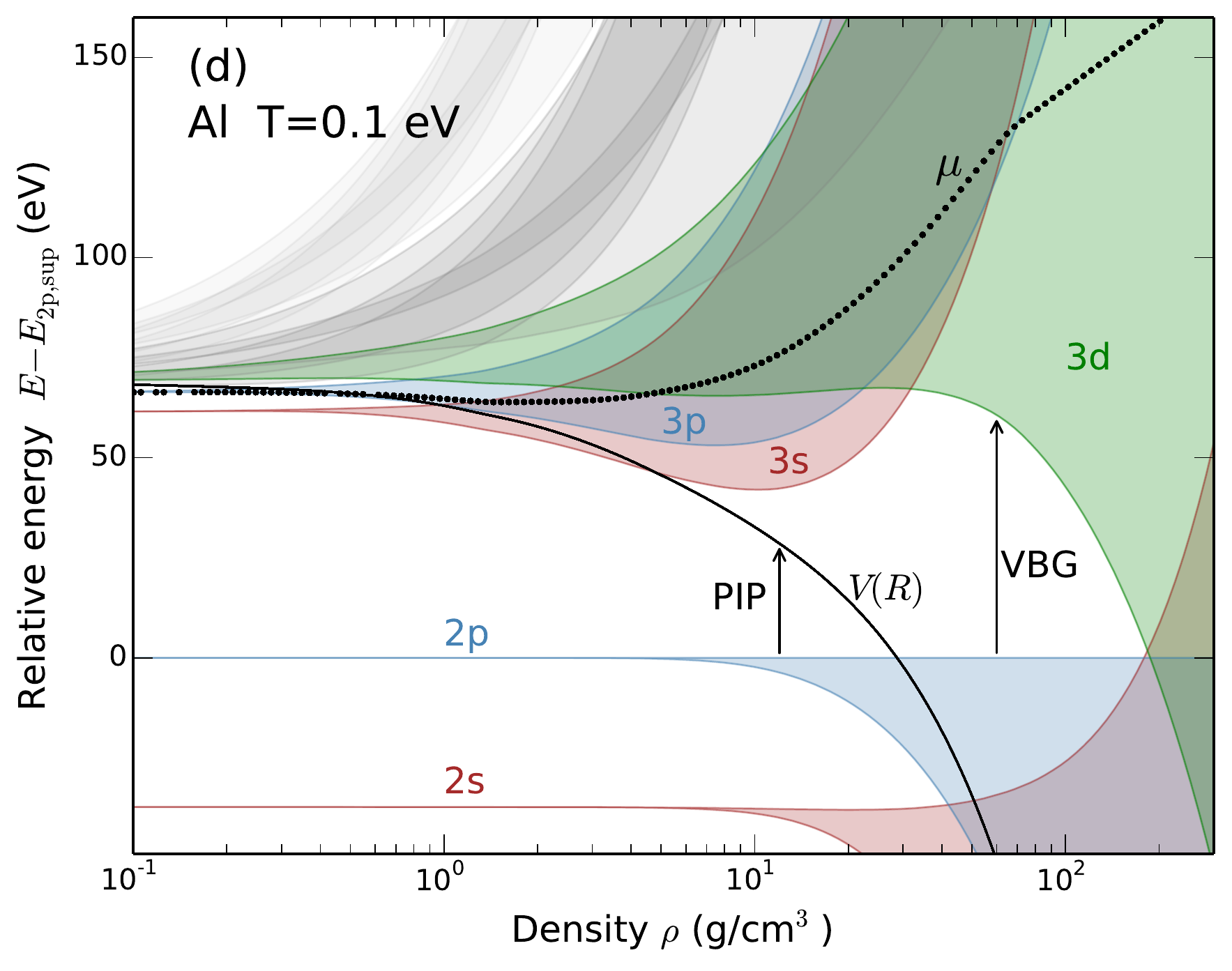} 
    \caption{Evolution of band energies computed with \Avion~as a function of density 
    for carbon (first column) and aluminum (second column), 
    for the temperatures $T=12.5$~eV (upper row) and $T=0.1$~eV (lower row). 
    In panels (a) and (b), energies are counted from the potential threshold $V(R)$;
    in (c) from the top of the 1s band for carbon; in (d) from the top of the 2p band for aluminum. 
    Lowest s, p and d bands are in red, blue and green colors respectively. 
    The black line is the potential threshold $V(R)$; 
    the black dots mark the chemical potential $\mu$ at computed points. 
    The valence band gap (VBG) and the pseudo-ionization potential (PIP) of the C 1s and Al 2p bands 
    are shown as vertical arrows. 
    In panel (c), an arrow indicates the carbon K-edge, which is in good agreement with 
    Hu's predictions~\cite{Hu:2017ab} for $T=1.3$~eV (blue dots).}  
    \label{fig:C_and_Al_AvIonBands}
\end{figure*}

As an illustration of our
approach, 
Fig.~\ref{fig:C_and_Al_AvIonBands}  shows
the evolution of band energies over a large density range for carbon and aluminum at 
temperatures $T=12.5$~eV and  $T=0.1$~eV. 
The reference energy in panel (a) and (b) is the potential threshold $V(R)$. 
In order to better visualize the VBG, energies are counted from the top of  
the 1s band for carbon in (c), and the top of the 2p band for aluminum in (d). 
At low density, bands below the threshold appear as sharp, localized bound states. With increasing density, they widen as they approach the potential threshold, which they 
eventually cross in a continuous way. 
Above a density of a few \gcc 
the bottom of the continuum  
is associated for carbon first with the 2s and later with the 2p band, while for Al, it is first the 3s and later the 3d band. 
The void region between the top of the valence band (1s for C, 2p for Al) and the continuum defines the VBG,  
as shown with arrows in the panels.

Our \Avion~model does not distinguish {\em bound} and {\em free} states. 
States rather evolve from being {\em localized} to {\em delocalized}. 
To compare with traditional AA and IPD models, we introduce a pseudo-ionization potential (PIP) in \Avion, which we define to be the difference between a state's energy and the threshold potential, $V(R)$, even though the PIP 
is not linked to any physical quantity in our approach. 
We show the PIP 
for the 1s state of C in panels (a) and (c) of Fig.~\ref{fig:C_and_Al_AvIonBands}, and for the 2p state of Al in (b) and (d). 
Both decrease rapidly with density. The VBG predicted by our \Avion~approach remains much larger because the region void of states extends well above the potential threshold. This explains discrepancies reported in Ref.~\citep{Driver:2018aa} between very small gaps in AA models and large VBGs in DFT-MD results,
as is demonstrated by further analysis in the following section.

In \Avion, we can predict the K-edge by subtracting the 1s state's energy from the chemical potential (or Fermi energy), 
as is illustrated for carbon in 
Fig.~\ref{fig:C_and_Al_AvIonBands}(c). In the same panel, we have included as blue dots the values computed from DFT simulations in Ref.~\cite{Hu:2017ab} for $T=1.3$~eV. The agreement with the K-edge inferred from our \Avion~approach is excellent. 
The increase of the K-edge energy emerges as a 'competition between continuum lowering and Fermi surface rising' 
\cite{Hu:2017ab,Iglesias:2018ab,Hu:2018aa}
when including the potential threshold as an intermediate step in the analysis. 
Still in \Avion~and in DFT, the K-edge energy can be determined directly. 

\begin{figure}
    \includegraphics[width=0.5\textwidth]{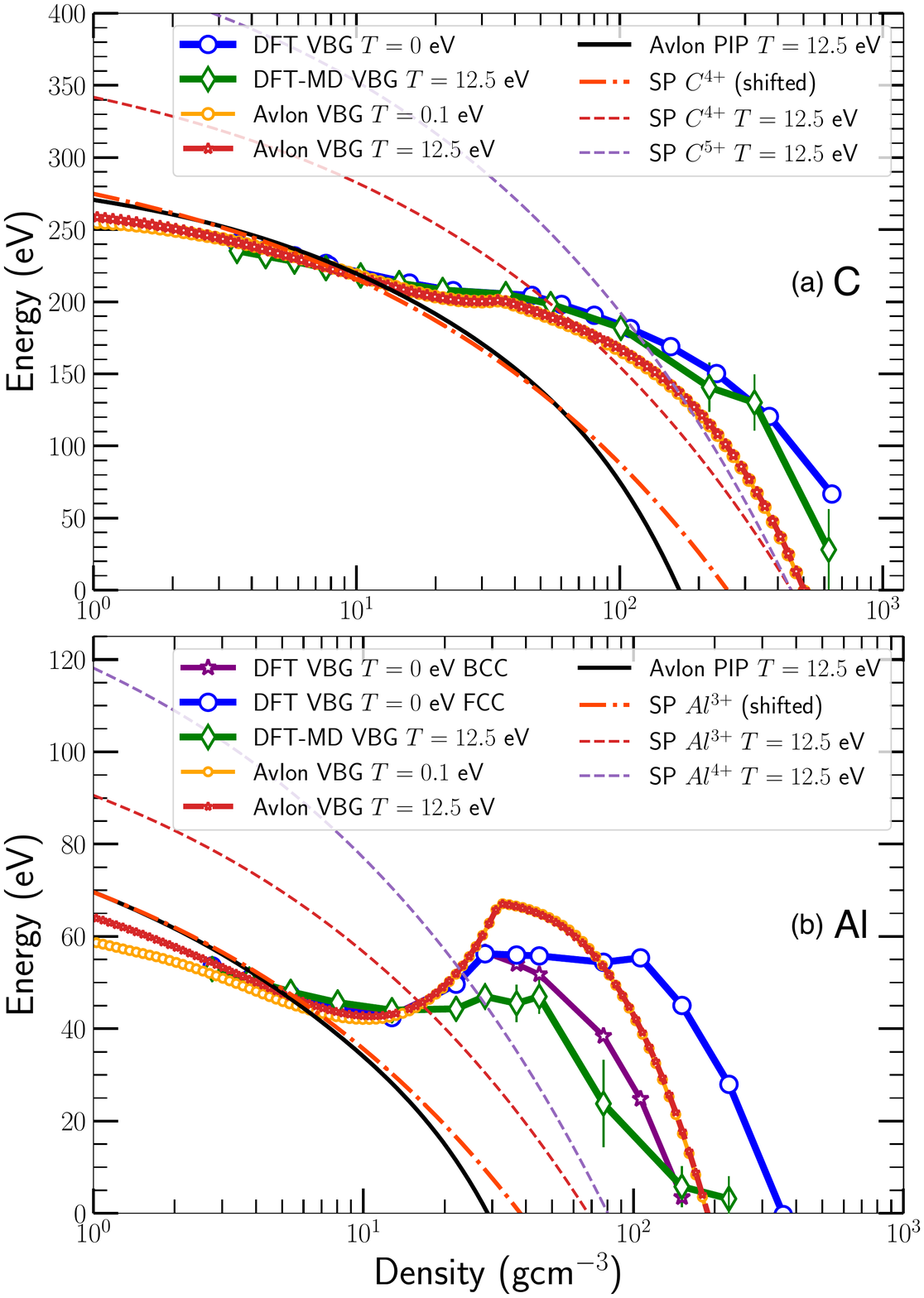}   
    \caption{Valence band gaps (VBG) of (a) carbon and (b) aluminum. We compare static DFT calculations of solids ($T=0$), DFT-MD simulations of liquids at $T=12.5$~eV, and 
    \Avion~results. The pseudo-IP (PIP) from \Avion~is compared with SP predictions for the IP of 
    various ionization states, as well as a shifted curve for the less ionized ion (dash-dotted). 
    The SP C$^{4+}$ curve in (a) has been shifted by $-67$~eV, and the SP Al$^{3+}$ curve in (b) by $-21$~eV.     }
    \label{fig:bg_c_and_al}
\end{figure}

\section{Comparison between methods} 

The DOS as obtained from \Avion~are compared in Fig.~\ref{fig:DOS_ALL} to DFT-MD results at $T=12.5$~eV and for the lowest densities. To ease comparison, the DOS have been convolved using a Gaussian distribution for the neutrality radii $R$ with a HWHM equal to $R_{\mathrm WS}/40$, which slightly broadens the valence peaks. As \Avion~is based on a quite simple prescription for the DOS shape inside a band, a perfect match is not expected. Nevertheless the model reproduces remarkably the DFT-MD results and their trends, especially since it requires a few seconds against hundreds of hours for DFT. The continuum edges and the valence band positions agree quite well for carbon, and for aluminum at the lowest density of 8.03~\gcc. 
For Al at 22.0~\gcc the agreement on the positions of both the 2s and 2p bands and the continuum edge can be made better 
by upshifting with $\sim10$~eV the \Avion~DOS (which amounts to a shift of the chemical potential). 

The position of the potential thresholds $V(R)$ are also drawn as vertical bars in the top of the panels. For carbon, the continuum starts just below the threshold at $3.51$~\gcc, and just above at $21.2$~\gcc. For aluminum, it is just below the threshold at $8.03$~\gcc. 
These situations would be hard to distinguish experimentally from the usual equality in traditional AA models.
In contrast, for aluminum at $22.0$~\gcc, the continuum edge starts some $40$~eV above the threshold. These results are easily visualized through inspection of the upper panels of Fig.~\ref{fig:C_and_Al_AvIonBands}.

In Fig.~\ref{fig:bg_c_and_al}, we compare results of DFT, \Avion, and SP methods over three orders of magnitude in density. For both $T=0$ conditions and fluid states at $T=12.5$~eV, we show results from DFT(-MD) simulations and \Avion~calculations. The latter use $T=0.1$~eV instead of $T=0$ to avoid numerical issues, but as a matter of fact the \Avion~results are fairly similar in these low temperature range except for a slight difference for Al at low density.

For carbon, \Avion~reproduces the DFT band gaps very well but a small deviation remains for the highest 
densities. \Avion~predicts a gap closure at a density of $\approx500$~\gcc 
while the gap in static DFT calculations of fcc carbon has not yet closed. DFT-MD simulations of liquid carbon predict gap closure at a density of $\approx620$~\gcc. (We define the point of gap closure in DFT-MD when the standard deviation of the distribution of gap values overlaps with zero.) 

For aluminum, the agreement is very good up to a density of around 20~\gcc. Then the \Avion~band gap increases to higher values than is predicted with DFT methods. However, the following decrease and eventual band gap closure at $\approx190$~\gcc occurs in a range compatible with DFT predictions. The difference in the latter may be traced back to the different crystal or fluid state in the DFT simulations, while the environment in \Avion~is always with spherical symmetry. Note that \Avion~shows a marked change of the band gap slope around 32~\gcc, which is linked to the switch from the 3s to the 3d band used to define the continuum edge (see Fig.~\ref{fig:C_and_Al_AvIonBands}(d)). This change in slope is also seen in our DFT simulations. A similar slope change, though less dramatic, occurs for carbon at around 35~\gcc due to the switch of the lowest continuum band from 2s to 2p.

In Fig.~\ref{fig:bg_c_and_al}, we also plot the PIP computed with \Avion~for $T=12.5 $ eV
(it is almost the same for $T=0.1$~eV on this density range.)
Its value goes to zero at already $\approx170$~\gcc for C and $\approx30$~\gcc  for Al. 
From the point of view of 
traditional AA models, this would imply that the involved bands have already merged with the continuum at these conditions. However, both our \Avion~and DFT results predict there is still a sizable VBG present under these conditions. 

It is useful to compare when \Avion~and SP models predict a given state to become pressure ionized. 
From around 1~\gcc up to the density where the valence band crosses the threshold, carbon has two localized electrons and aluminum has ten. We hence plot in Fig.~\ref{fig:bg_c_and_al} the modified IP,  $E^{0}-\Delta_{SP}$, for the 1s state of C$^{4+}$ and C$^{5+}$, and for the 2p state of Al$^{3+}$ and Al$^{4+}$, where $E^{0}$ is the isolated ion spectroscopic value and $\Delta_{SP}$ the IPD predicted by the SP model. It shows little resemblance to the VBG curves computed with \Avion~and DFT. SP models predict pressure ionization to occur at lower densities, at values in between the point where \Avion's PIP goes to zero and \Avion~and DFT predict the VBG to close.

It should be noted that AA models and DFT simulations involve approximations which do not allow reproduction of 
the measured energies of isolated ions, which are input parameters in SP models. By shifting SP curves for C$^{4+}$ and Al$^{3+}$, i.e. adjusting $E^{0}$, it is possible to match the evolution of \Avion's PIP over most of the density range
(see lines labelled "shifted" in Fig.\ref{fig:bg_c_and_al}.) 
In these degenerate conditions, \Avion~and SP models make similar predictions for screening effects and $\Delta_{SP}$ reduces to the ion sphere expression, $3ze^2/(2R)$, where $z=4$ for C and $z=3$ for Al. 
The residual discrepancy for the densities where both curves go to zero may be explained by noticing that the PIP is counted from the top of the valence band; the SP model rather refers to an eigenstate that would be computed with the condition $P_{\varepsilon\ell}(\infty)=0$. Hence it is situated in the band below its top and is ionized at a higher density.

\section{Conclusion} 

In this paper, 
we have developed a novel type of AA approach,  \Avion, to obtain consistency with much more computationally demanding many-body DFT simulations. We studied solid and liquid carbon and aluminum over three orders of density. With our \Avion~method, we find excellent agreement with DFT predictions for the continuum edge shift below the Fermi energy, the carbon K-edge increase and the valence band gaps. Traditional AA models on the other hand predict 
valence bands to enter the continuum
at far too low densities. Still, our \Avion~results agree with a continuum lowering predicted by SP model up to a few times solid density. But at higher densities, drastic errors in the SP models become apparent as the continuum band shifts away from the potential threshold that is incorrectly equated with the beginning of the continuum band in the SP model. 

Such predictions may be tested in experiment at high energy laser facilities like the National Ignition Facility or Laser M\'egajoule (LMJ) \cite{LMJ} that are capable of producing compressed matter of Gbar pressures~\cite{Kritcher2020} and multiple times solid density~\cite{Smith2014} and may be 
combined 
with x-ray diffraction~\cite{Lazicki2021} and x-ray Thomson scattering~\cite{Chapman2014}.

Since our \Avion~approach is as efficient as traditional AA models and can deliver DFT compatible predictions in only a few CPU seconds, it allows to fit and interpret experimental data and infer the temperature and density without resorting to SP models. The capability of \Avion~will make joint experimental-theoretical explorations of WDM more efficient.


\begin{acknowledgments}
  GM acknowledges support from the Programme National de Physique Stellaire (PNPS) of CNRS/INSU, MB from the Center of Advanced Systems Understanding (CASUS), FS from the European Union through a Marie Sk\l odowska-Curie action (grant 750901) and BM from the National Science Foundation-Department of Energy (DOE) partnership for plasma science
  and engineering (grant DE-SC0016248) and by the DOE-National Nuclear
  Security Administration (grant DE-NA0003842). The DFT and DFT-MD simulations were performed on a Bull Cluster at the Center for Information Services and High Performace Computing (ZIH) at
Technische Universit\"at Dresden.
\end{acknowledgments}

%


\end{document}